\documentclass[twocolumn,english,prl,groupedaddress,floatfix]{revtex4}
\usepackage[T1]{fontenc}
\usepackage[latin9]{inputenc}
\usepackage{graphicx}

\makeatletter

\usepackage{esint}

\makeatletter
\@ifundefined{textcolor}{}
{%
 \definecolor{BLACK}{gray}{0}
 \definecolor{WHITE}{gray}{1}
 \definecolor{RED}{rgb}{1,0,0}
 \definecolor{GREEN}{rgb}{0,1,0}
 \definecolor{BLUE}{rgb}{0,0,1}
 \definecolor{CYAN}{cmyk}{1,0,0,0}
 \definecolor{MAGENTA}{cmyk}{0,1,0,0}
 \definecolor{YELLOW}{cmyk}{0,0,1,0}
 }

\usepackage{bm}
\usepackage{babel}

\makeatletter
\@ifundefined{textcolor}{}
{
 \definecolor{BLACK}{gray}{0}
 \definecolor{WHITE}{gray}{1}
 \definecolor{RED}{rgb}{1,0,0}
 \definecolor{GREEN}{rgb}{0,1,0}
 \definecolor{BLUE}{rgb}{0,0,1}
 \definecolor{CYAN}{cmyk}{1,0,0,0}
 \definecolor{MAGENTA}{cmyk}{0,1,0,0}
 \definecolor{YELLOW}{cmyk}{0,0,1,0}
 }
\makeatother

\makeatother

\usepackage{babel}

\makeatother

\usepackage{babel}

\begin{document}

\title{Scaling of nascent nodes in extended s-wave superconductors}

\author{Rafael M. Fernandes and J\"org Schmalian}

\affiliation{Ames Laboratory and Department of Physics and Astronomy, Iowa State
Univ., Ames, IA 50011, USA }

\date{\today }

\begin{abstract}
We analyze the low-energy properties of superconductors near the onset
of accidental nodes, i.e. zeroes of the gap function not enforced
by symmetry. The existence of such nodes has been motivated by recent
experiments suggesting a transition between nodeless and nodal superconductivity
in iron-based compounds. We find that the low-temperature behavior
of the penetration depth, the specific heat, and the NMR-NQR spin-lattice
relaxation rate are determined by the scaling properties of a quantum
critical point associated with the nascent nodes. Although the power-law
exponents are insensitive to weak short-range electronic interactions,
they can be significantly altered by the curvature of the Fermi surface
or by the three-dimensional character of the gap. Consequently, the
behavior of macroscopic quantities near the onset of nodes can be
used as a criterion to determine the nodal structure of the gap function. 
\end{abstract}
\maketitle
After three years of their discovery, the symmetry of the superconducting
state of the iron-based compounds is still under intense debate \cite{Johnston10}.
Although many early probes suggested a fully gapped state, recent
experiments have found strong evidence for the presence of gapless
quasi-particle excitations, suggesting the presence of nodes in at
least some regions of the phase diagram \cite{Martin10,Martin10_2,Tanatar10,Reid10,Jang,Wu10,Fischer10,Hashimoto10,Reid11}.
Interestingly, in overdoped $\mathrm{Ba}\left(\mathrm{Fe}_{1-x}TM_{x}\right)_{2}\mathrm{As_{2}}$
materials, with $TM=\mathrm{Co,Ni},\mathrm{Pd}$, penetration depth
\cite{Martin10,Martin10_2}, specific heat \cite{Jang}, and thermal
conductivity measurements \cite{Tanatar10,Reid10} indicate a transition
from a nodeless to an anisotropic nodal state, that does not seem
to be related to changes in the Fermi surface (FS) topology. In underdoped
$\left(\mathrm{Ba}_{1-x}\mathrm{K}_{x}\right)\mathrm{Fe_{2}As_{2}}$
compounds, where superconductivity (SC) may coexist with the magnetically
ordered phase, recent thermal conductivity and penetration depth measurements
also show a doping-induced onset of nodal quasi-particles \cite{Reid11,Kim11}.

One possible scenario to explain these nodeless-to-nodal transitions
is a change in symmetry from a SC state with finite gap everywhere
on the FS, e.g. an isotropic $s$-wave pairing state, to a state with
symmetry-enforced nodes, such as a $d$-wave state. In this case,
a line of finite temperature phase transitions would separate the
nodal and nodeless SC states. Indeed, for a range of band structure
parameters, some theoretical models find nearly degenerate $s$-wave
and $d$-wave solutions of the gap equations associated with purely
electronic pairing interactions \cite{Graser_degenaracy,Maiti_degeneracy}.
Another scenario, not associated with a finite-temperature phase transition,
is the emergence of accidental nodes, i.e. zeroes of the gap function
at positions of the FS unrelated to a particular representation of
the tetragonal symmetry group. In the iron pnictides, one frequently
discussed gap function $\Delta\left(\mathbf{k}\right)$ is the extended
$s$-wave, with different signs along different FS sheets \cite{Mazin08,Kuroki08,Chubukov08}.
The location of its zeroes in momentum space is determined by energetic
arguments rather than symmetry requirements. Therefore, changing the
size of the FS sheets or modifying quantitative details of the pairing
interaction can cause such a zero to touch and eventually cross one
of the FS pockets, leading to a nodal state \cite{nodes_Chubukov,nodes_Maier,nodes_Sknepnek,nodes_wang,nodes_platt,nodes_mishra,nodes_Kemper,nodes_eremin,nodes_Maiti,nodes_Thomale,nodes_Mishra2}.

\begin{figure}
\begin{centering}
\includegraphics[width=0.6\columnwidth]{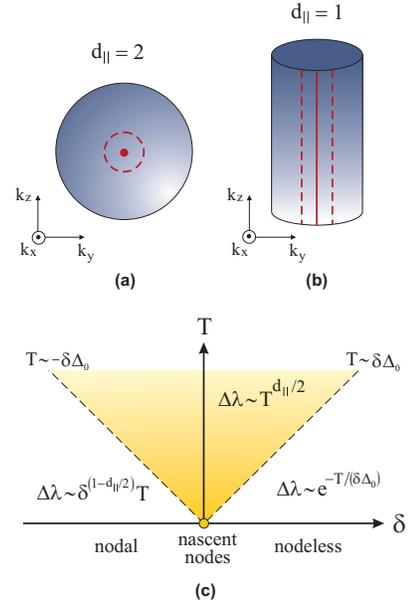} 
\par\end{centering}

\caption{Accidental nodes in spherical (a) and cylindrical (b) Fermi surfaces.
The point and the full line denote the $\delta=0$ nascent node in
each case, whereas the dashed lines denote the line nodes that develop
for $\delta<0$. $d_{\parallel}$ is the number of directions on the
FS with curvature. In (c) we present the $T$-$\delta$ phase diagram
with the crossover regime (shaded yellow region between the dashed
lines) and the scaling properties of the variation of the penetration
depth $\Delta\lambda$.}

\end{figure}

At finite temperatures, the onset of accidental nodes does not imply
a phase transition. However, it is characterized by a quantum critical
point (QCP) that separates a regime with finite-ranged quasi-particle
excitations, in the fully gapped state, from a regime with algebraically
decaying excitations, in the nodal state. Thus, general crossover
arguments of QCP should be useful to characterize the behavior of
macroscopic quantities near the onset of accidental nodes. Consider,
for example, the variation of the penetration depth $\Delta\lambda=\lambda\left(T\right)-\lambda\left(0\right)$
(we only consider the clean case here): in the gapped state one expects
activated behavior, $\Delta\lambda\propto\exp\left(-\Delta/T\right)$,
whereas in the case of line nodes holds $\Delta\lambda=AT$, with
constant $A$. Naively, one expects that the transition between the
two regimes is characterized by a power-law $\Delta\lambda=A'T^{n}$
with exponent $1<n<\infty$ intermediate between the two regimes.
For instance, if the accidental nodes first touch the FS at a single
point, one would expect to observe $n=2$, characteristic of systems
with symmetry-enforced point nodes (see Ref.\cite{Sigrist} for a
review).

In this Brief Report we analyze the crossover behavior near the onset
of accidental nodes and how it depends on the shape of the Fermi surface.
Among others, our results demonstrate that even if nascent nodes originate
around a single point, they behave qualitatively different from symmetry-enforced
point nodes. Our main results are summarized in Fig.1: in the case
of a spherical FS, at the onset of accidental nodes, we obtain a linear-in-$T$
penetration depth ($n=1$), but with a coefficient that is precisely
half of the value that one finds for line nodes, $A'=A/2$. Consequently,
after the onset of accidental nodes, $\Delta\lambda$ displays a crossover
between two linear-in-$T$ regimes with different slopes. Similar
behaviors hold for other quantities, such as the specific heat $C$
and the NMR-NQR spin-lattice relaxation rate $T_{1}$. In the case
of a perfectly cylindrical Fermi surface we find, in agreement with
earlier calculations \cite{Stanev10}, that $n=1/2$, i.e. a behavior
more singular even than for line nodes. Denoting by $d_{\parallel}$
the number of directions on the FS with curvature ($d_{\parallel}=2$
for a spherical FS and $d_{\parallel}=1$ for a perfectly cylindrical
FS), our results for the scaling behavior are expressed as $\Delta\lambda\propto T^{d_{\parallel}/2}$,
$C\propto T^{1+d_{\parallel}/2}$, and $T_{1}^{-1}\propto T^{1+d_{\parallel}}$.
In order for the more exotic regime of cylindrical FS ($d_{\parallel}=1$)
to be observed, the energy associated with the FS curvature must be
negligible on the scale of the SC gap $\Delta_{0}\simeq10$ meV, which
is a very restrictive condition. Yet, this sensitivity with respect
to the shape of the FS might enable one to distinguish between distinct
nodal structures of the SC gap. In what follows, we will derive these
results and show that they are robust against weak short range electron-electron
interactions.

To describe the onset of accidental nodes we consider the angle-dependent
gap function on the FS: \begin{equation}
\frac{\Delta\left(\theta\right)}{\Delta_{0}}=\frac{1+\delta}{2}-\frac{1-\delta}{2}\cos\left(2\theta\right).\label{gap}\end{equation}

Here $\Delta_{0}>0$ is the maximum amplitude of the gap, $\theta$
is the angle measured relative to the center of the FS, and $\delta\ $is
the tuning parameter. For $\delta>0$ the superconductor is fully
gapped with minimum gap amplitude $\Delta_{0}\delta$. For $\delta<0$,
the gap function changes sign at nodes located at angles not fixed
by symmetry, $\theta_{0\pm}=\pm\frac{1}{2}\arccos\left(\frac{1+\delta}{1-\delta}\right)$,
and its most negative value is $\Delta_{0}\delta<0$ . The special
case $\delta=0$ corresponds to the onset of nodes at $\theta_{0}=0$
and $\pi$. Our results do not depend on the specific form of $\Delta\left(\theta\right)$
given in Eq.\ref{gap}; rather, they apply generally to systems where,
close to the nodes, the gap can be written as $\Delta\left(\theta\right)\simeq\Delta_{0}\left(\theta+\theta_{0+}\right)\left(\theta-\theta_{0-}\right)$.
In this language, the QCP is characterized by the merging of the two
nodes, giving rise to a quadratic dispersion $\Delta\left(\theta\right)\simeq\Delta_{0}\left(\theta-\theta_{0}\right)^{2}$.
In the context of the iron pnictides, $\theta$ in Eq.\ref{gap} could
be an angle relative to the center of the electron pocket. If one
considers instead accidental nodes in the hole pocket, one simply
replaces $\cos\left(2\theta\right)$ by $\cos\left(4\theta\right)$,
which does not change the results obtained here.

The low temperature behavior of numerous observables \cite{Hirschfeld,Sigrist}
is determined by the density of states of Bogoliubov quasi-particles:
\begin{equation}
N\left(\omega\right)=N_{0}\int d\varepsilon d\Omega\sum_{l=\pm1}\delta\left(\omega-l\sqrt{\varepsilon^{2}+\Delta\left(\theta\right)^{2}}\right),\label{density_states}\end{equation}
 where $N_{0}$ is the normal-state density of states at the Fermi
level and $d\Omega$ refers to the angular integration. With $N\left(\omega\right)$,
it is straightforward to obtain the temperature dependence of the
penetration depth, the specific heat, and the NMR-NQR spin-lattice
relaxation rate. Denoting the Fermi function by $f\left(\omega\right)=\left(e^{\beta\omega}+1\right)^{-1}$,
with $\beta=\left(k_{B}T\right)^{-1}$, one obtains:

\begin{eqnarray}
\Delta\lambda & \propto & \int_{0}^{\infty}d\omega N\left(\omega\right)\left(-\frac{\partial f\left(\omega\right)}{\partial\omega}\right),\nonumber \\
C & \propto & \int_{0}^{\infty}d\omega N\left(\omega\right)\beta\omega^{2}\left(-\frac{\partial f\left(\omega\right)}{\partial\omega}\right),\nonumber \\
\frac{1}{T_{1}T} & \propto & \int_{0}^{\infty}d\omega N\left(\omega\right)^{2}\left(-\frac{\partial f\left(\omega\right)}{\partial\omega}\right).\label{observables}\end{eqnarray}

Band structure calculations as well as ARPES measurements reveal that
the FS of the iron pnictides are anisotropic, similar in shape to
warped cylinders \cite{Kondo10}. Although small, this curvature of
the FS becomes very relevant for the low energy properties of interest
here. Thus, close to the accidental nodes, the FS is similar in shape
to an ellipsoid which, upon rescaling, can be described by a sphere,
yielding $d\Omega=\left(4\pi\right)^{-1}\sin\theta d\theta d\varphi$
in Eq. \ref{density_states}. In this case the accidental line node
becomes elliptical in shape, starting at an isolated point (see Fig.1a).
The resulting density of states is given by:

\begin{equation}
N\left(\omega\right)=\frac{N_{0}}{\sqrt{1-\delta}}\int_{\max\left(\delta\Delta_{0},-\omega\right)}^{\min\left(\Delta_{0},\omega\right)}\frac{\omega}{\sqrt{\omega^{2}-\Delta^{2}}}\frac{d\Delta}{\sqrt{\Delta_{0}-\Delta}}\label{density_states_2}\end{equation}

In Fig. 2, we show $N\left(\omega\right)$ for $\delta$ larger, smaller
and equal to zero. At low energies $\left\vert \omega\right\vert \ll\Delta_{0}$
and near the onset of nodes (i.e. for $\left\vert \delta\right\vert \ll1$),
we obtain: \begin{equation}
\frac{N\left(\omega\right)}{N_{0}}=\left\{ \begin{array}{lc}
\frac{\left|\omega\right|\mathrm{arcsec}\left(\frac{\left|\omega\right|}{\delta\Delta_{0}}\right)}{\Delta_{0}\sqrt{1-\delta}}\:\Theta\left(\left|\omega\right|-\delta\Delta_{0}\right) & \:\delta>0\\
\frac{\pi}{2}\frac{\left\vert \omega\right\vert }{\Delta_{0}} & \:\delta=0\\
\frac{\pi}{\sqrt{1-\delta}}\frac{\left\vert \omega\right\vert }{\Delta_{0}}\: g\left(\frac{\omega}{\left\vert \delta\right\vert \Delta_{0}}\right) & \:\delta<0\end{array}\right.\label{sphere}\end{equation}
 where $g\left(x\right)=1$ for $\left\vert x\right\vert <1$ and
$g\left(\omega\right)=\ \frac{1}{2}+\frac{1}{\pi}\mathrm{arccot}\sqrt{x^{2}-1}$
for $\left\vert x\right\vert >1$. Notice that the nascent point nodes
at $\delta=0$ behave differently than the symmetry enforced point
nodes occurring in the polar state of triplet superconductors \cite{Hirschfeld,Sigrist}.
In the latter, the gap magnitude near the node varies as $\left\vert \theta-\theta_{0}\right\vert $,
as opposed to $\left(\theta-\theta_{0}\right)^{2}$ discussed here,
giving rise to $N\left(\omega\right)\propto\omega^{2}$ instead of
the $N\left(\omega\right)\propto\omega$ found here. A similar result
was discussed in Ref. \cite{Graf96}, in the context of hybrid pairing
states of uniaxial superconductors.

\begin{figure}
\begin{centering}
\includegraphics[width=0.7\columnwidth]{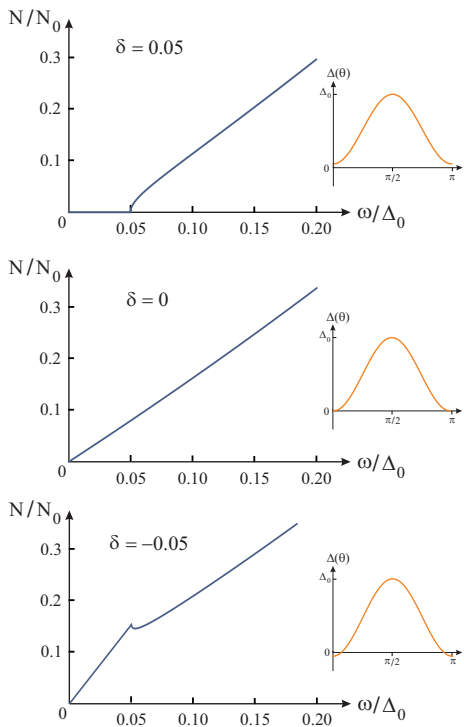} 
\par\end{centering}

\caption{Density of states $N$ as function of energy $\omega$ in the case
of a spherical Fermi surface for different values of $\delta$. The
insets show the dependence of the gap $\Delta$ on the polar angle
$\theta$.}

\end{figure}

Because of the shallow rise of the gap near the nascent nodes, the
slope of the density of states for $\delta=0$ is only half of the
value for $\delta<0$, where the gap actually changes sign. Consequently,
for $\delta<0$ a crossover takes place between two regimes that are
linear in $\omega$, with a low-energy slope twice as large as the
higher-energy slope. In the fully gapped regime, $\delta>0$, the
density of states vanishes as $N_{0}\sqrt{\frac{2\delta}{1-\delta}}\sqrt{\frac{\left\vert \omega\right\vert -\delta\Delta_{0}}{\Delta_{0}}}$
near the threshold, but soon recovers the linear critical behavior
at higher energies. This $T=0$ crossover regime is also manifested
at finite temperatures, as we show in Fig. 3, where the $T$-dependence
of the penetration depth is displayed for different values of the
tuning parameter $\delta$. To probe the low-energy behavior of $N\left(\omega\right)$,
low temperatures are required\textbf{. }Therefore, the observation
of this crossover, or of a similar crossover in the slope of $C/T$
or $1/T^{2}T_{1}$, would be a strong evidence for accidental nodes
and thus extended $s$-wave pairing.

\begin{figure}
\begin{centering}
\includegraphics[width=0.85\columnwidth]{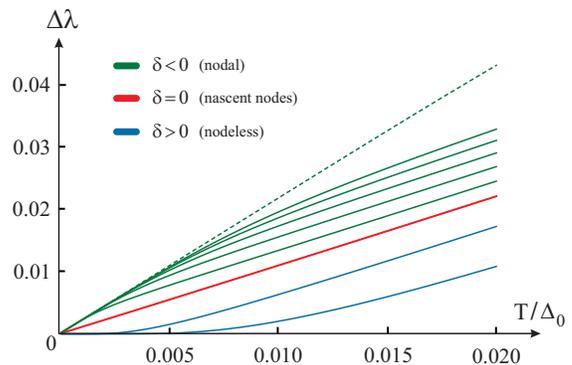} 
\par\end{centering}

\caption{Variation of the penetration depth $\Delta\lambda\equiv\lambda\left(T\right)-\lambda\left(0\right)$
as function of temperature $T$ for several values of $\delta$. Notice
in all curves the crossover at higher temperatures to the linear-in-$T$
dependence associated with the nascent nodes (solid red line). At
low temperatures, $\Delta\lambda$ decays exponentially for the fully
gapped case (solid blue line) and linearly for the nodal case, but
with a different slope (dashed green line). Here we used the values
$\delta=0.025$, $0.01$, $0$, $-0.005$, $-0.01$, $-0.015$, $-0.02$,
and $-0.025$. }

\end{figure}

For completeness, we also discuss the regime of perfectly cylindrical
FS. In this case the angular integration becomes $d\Omega=\left(2\pi\right)^{-1}d\theta$
and two nearly parallel line nodes emerge from a single line of nascent
nodes with $\Delta\left(\theta,k_{z}\right)\simeq\Delta_{0}\left(\theta-\theta_{0}\right)^{2}$
independent of $k_{z}$ (see Fig. 1b). The low-energy density of states
is then given by\begin{equation}
\frac{N\left(\omega\right)}{N_{0}}=\left\{ \begin{array}{lc}
\left(\frac{2\left|\omega\right|}{\Delta_{0}}\right)^{1/2}\Theta\left(\left|\omega\right|-\delta\Delta_{0}\right) & \delta>0\\
2\left(2\pi\right)^{-3/2}\Gamma\left(\frac{1}{4}\right)^{2}\left(\frac{\left\vert \omega\right\vert }{\Delta_{0}}\right)^{1/2} & \delta=0\\
2\left\vert \delta\right\vert ^{-1/2}\frac{\left\vert \omega\right\vert }{\Delta_{0}}\:\widetilde{g}\left(\frac{\left\vert \omega\right\vert }{\delta\Delta_{0}}\right) & \delta<0\end{array}\right.\label{cyl}\end{equation}
 with $\widetilde{g}\left(x\right)=1$ for $\left\vert x\right\vert \ll1$,
$\widetilde{g}\left(x\right)=\left(2\pi\right)^{-3/2}\Gamma\left(\frac{1}{4}\right)^{2}\left\vert x\right\vert ^{-1/2}$
for $1\ll\left\vert x\right\vert \ll\left\vert \delta\right\vert ^{-1}$,
and a logarithmically diverging $\widetilde{g}\left(x\right)=\sqrt{2\left\vert \delta\right\vert /\pi^{2}}\log\left(\frac{1}{\left\vert x\right\vert -1}\right)$
close to $x=1$. The surprising aspect of this result is that the
power in $N\left(\omega\right)$ at the QCP is lower even than the
case of line nodes, a result that was obtained earlier in Ref. \cite{Stanev10}.
Thus, right at the transition to a fully gapped regime there are more
low energy quasi-particle excitations compared to the nodal regime,
again a consequence of the shallow rise of the gap for arbitrary $k_{z}$.
Since the actual FS is not perfectly cylindrical, one expects a crossover
to the linear in $\omega$ behavior for a spherical FS, Eq. \ref{sphere}.\textbf{
}This crossover will depend on details of the $k_{z}$-dispersions
of the FS (as observed in \cite{Kondo10}) and of the SC gap (as observed
in \cite{Arpes01,Arpes02}) - in particular, on how the curvature
of the FS changes as function of energy when compared to the changes
in the SC gap. In Ref. \cite{Reid10}, Reid \emph{et al.} suggest
the two nodal structures associated with Eqs. \ref{sphere} and \ref{cyl}
as the main candidates to explain their thermal conductivity measurements
in overdoped $\mathrm{Ba}\left(\mathrm{Fe}_{1-x}\mathrm{Co}_{x}\right)_{2}\mathrm{As_{2}}$.
The different crossover behaviors of several macroscopic quantities
that results from Eqs. \ref{sphere} and \ref{cyl} provide an important
criterion to decide between the two scenarios.

The results above can also be obtained from scaling arguments of the
low-energy action associated with the QCP. The free fermion action
for $\delta=0$ is \begin{equation}
S_{0}=\int_{k}\psi_{k}^{\dagger}\left(\begin{array}{cc}
i\omega-vk_{\perp} & \alpha k_{\parallel}^{2}\\
\alpha k_{\parallel}^{2} & i\omega+vk_{\perp}\end{array}\right)\psi_{k}\end{equation}
 where we used the Nambu spinors $\psi_{k}=\left(c_{k\uparrow},c_{-k\downarrow}^{\dagger}\right)$
to denote the nodal quasi-particles. Here $k=\left(k_{\perp},\mathbf{k}_{\parallel},i\omega\right)$
stands for momenta perpendicular and parallel to the FS measured from
the location of the nascent node, with $\int_{k}=T\sum_{n}\int dk_{\perp}d^{d_{\parallel}}k_{\parallel}$.
Upon scaling $k_{\perp}\rightarrow bk_{\perp}$, $k_{\parallel}\rightarrow b^{1/2}k_{\parallel}$
and $\omega\rightarrow b\omega$, the action becomes scale invariant
if one uses the rescaled field $\psi_{k}=b^{\kappa}\psi_{k^{\prime}}^{\prime}$
with $\kappa=\left(6+d_{\parallel}\right)/4$. Thus, we find the scaling
dimension of short range electron-electron interactions to be $-d_{\parallel}/4$.
Since this scaling dimension is negative, weak short range electronic
interactions are irrelevant in the sense of the renormalization group
theory and should not lead to an instability. It remains to be analyzed
the effects of long range interactions, such as quantum phase fluctuations.

Since the scaling dimension of the tuning parameter $\delta$ is unity,
we obtain the low-energy scaling form for the density of states: \begin{equation}
N\left(\omega,\delta\right)=b^{-d_{\parallel}/2}N\left(b\omega,b\delta\right).\end{equation}

For $\delta=0$, we can fix the arbitrary scaling parameter $b=\omega^{-1}$
which yields $N\left(\omega,0\right)\propto\omega^{d_{\parallel}/2}$.
For finite $\delta<0$, it holds that the slope $A\left(\delta\right)=\left.dN\left(\omega,\delta\right)/d\omega\right\vert _{\omega=0}$
behaves as $A\left(\delta\right)\propto\left\vert \delta\right\vert ^{1-d_{\parallel}/2}$.
These results are fully consistent with the explicit analysis presented
above.

In summary, we analyzed the scaling behavior near the onset of accidental
nodes in extended $s$-wave superconductors, and how it depends on
the curvature of the FS. The low-$T$ properties of the system can
be analyzed in a simple scaling theory that is governed by a QCP separating
a nodal and a nodeless region, without symmetry breaking. A somewhat
similar behavior was proposed in extreme type-II superconductors,
where the spectrum of Landau levels becomes gapless beyond a threshold
magnetic field \cite{Zlatkofield}. Our theory predicts that the low-temperature
slope of the linear quasi-particle density of states at the onset
of accidental nodes is reduced by a factor of $1/2$ compared to the
regime with well established nodes. This leads to an experimentally
observable crossover in a number of physical observables, allowing
one to distinguish whether nodes are accidental or symmetry enforced.

We thank A. V. Chubukov, R. T. Gordon, S. A. Kivelson, S. Maiti, R.
Prozorov, J.-Ph. Reid, L. Taillefer, and M. Tanatar for useful discussions.
Research at Ames Lab was supported by the U.S. DOE, Office of Basic
Energy Sciences, Materials Sciences and Engineering Division.


\begin{thebibliography}{10}
\bibitem{Johnston10} D. C. Johnston, Adv. Physics \textbf{59}, 803
(2010).

\bibitem{Martin10} C. Martin \emph{et al.}, Phys. Rev. B \textbf{81},
060505(R) (2010)

\bibitem{Martin10_2} C. Martin \emph{et al.},Supercond. Sci. Technol.
\textbf{23} 065022 (2010).

\bibitem{Tanatar10} M. Tanatar \emph{et al.}, Phys. Rev. Lett. \textbf{104},
067002 (2010).

\bibitem{Reid10} J.-Ph. Reid \emph{et al.}, Phys. Rev. B \textbf{82},
064501 (2010).

\bibitem{Jang} D.-J. Jang \emph{et al.}, New J. Phys. \textbf{13}
023036 (2011).

\bibitem{Wu10} D. Wu \emph{et al.}, Phys. Rev. B \textbf{82}, 184527
(2010).

\bibitem{Fischer10} T. Fischer \emph{et al.}, Phys. Rev. B \textbf{82},
224507 (2010).

\bibitem{Hashimoto10} K. Hashimoto \emph{et al.}, Phys. Rev. B \textbf{81},
220501(R) (2010).

\bibitem{Reid11} J.-Ph. Reid \emph{et al.}, arXiv:1105.2232 (2011).

\bibitem{Kim11} H. Kim \emph{et al}., arXiv:1105.2265 (2011).

\bibitem{Graser_degenaracy} S. Graser, T. A. Maier, P. J. Hirschfeld,
and D. J. Scalapino, New J. Phys. \textbf{11} 025016 (2009).

\bibitem{Maiti_degeneracy} S. Maiti \emph{et al}., arXiv:1104.1814
(2011).

\bibitem{Mazin08} I. I. Mazin, D. J. Singh, M. D. Johannes, and M.
H. Du, Phys. Rev. Lett. \textbf{101}, 057003 (2008).

\bibitem{Kuroki08} K. Kuroki \emph{et al}., Phys. Rev. Lett. \textbf{101},
087004 (2008).

\bibitem{Chubukov08} A. V. Chubukov, D. V. Efremov, and I. Eremin,
Phys. Rev. B \textbf{78}, 134512 (2008).

\bibitem{Cvetkovic09} V. Cvetkovic and Z. Tesanovic, EPL \textbf{85},
37002 (2009).

\bibitem{nodes_Chubukov} A. V. Chubukov, M. G. Vavilov, and A. B.
Vorontsov, Phys. Rev. B 80, 140515(R) (2009).

\bibitem{nodes_Maier} T. A. Maier, S. Graser, D. J. Scalapino, and
P. J. Hirschfeld, Phys. Rev. B \textbf{79}, 224510 (2009).

\bibitem{nodes_Sknepnek} R. Sknepnek, G. Samolyuk, Y. Lee, and J.
Schmalian, Phys. Rev. B \textbf{79}, 054511 (2009).

\bibitem{nodes_wang} F. Wang \emph{et al}., Phys. Rev. Lett. \textbf{102},
047005 (2009).

\bibitem{nodes_platt} C. Platt, C. Honerkamp, and W. Hanke, New J.
Phys. \textbf{11} 055058 (2009).

\bibitem{nodes_mishra} V. Mishra, A. B. Vorontsov, P. J. Hirschfeld,
and I. Vekhter, Phys. Rev. B \textbf{80}, 224525 (2009).

\bibitem{nodes_Kemper} A. F. Kemper \emph{et al.} New J. Phys. \textbf{12}
073030 (2010).

\bibitem{nodes_eremin} A. V. Chubukov and I. Eremin, Phys. Rev. B
\textbf{82}, 060504(R) (2010).

\bibitem{nodes_Maiti} S. Maiti and A. V. Chubukov, Phys. Rev. B \textbf{82},
214515 (2010).

\bibitem{nodes_Thomale} R. Thomale, C. Platt, W. Hanke, and B. A.
Bernevig, Phys. Rev. Lett. \textbf{106}, 187003 (2011).

\bibitem{nodes_Mishra2} V. Mishra, S. Graser, and P. J. Hirschfeld,
arXiv:1101.5699 (2011).

\bibitem{Sigrist} M. Sigrist and K. Ueda, Rev. Mod. Phys. \textbf{63},
239 (1991).

\bibitem{Stanev10} V. Stanev, B. S. Alexandrov, P. Nikolic, Z. Tesanovic,
arXiv:1006.0447 (2010).

\bibitem{Hirschfeld} P. J. Hirschfeld, P. W\"olfle, and D. Einzel,
Phys. Rev. B \textbf{37}, 83 (1988).

\bibitem{Kondo10} T. Kondo \emph{et al.}, Phys. Rev. B \textbf{81},
060507(R) (2010).

\bibitem{Graf96} M. J. Graf \emph{et al.}, Phys. Rev. B \textbf{53},
15147 (1996).

\bibitem{Arpes01} Y.M. Xu \emph{et al}., Nature Physics \textbf{7},
198 (2011).

\bibitem{Arpes02} Y. Zhang \emph{et al}., Phys. Rev. Lett. \textbf{105},
117003 (2010)

\bibitem{Zlatkofield} Z. Tesanovic and P. D. Sacramento, Phys. Rev.
Lett. \textbf{80}, 1521 (1998). 
\end{thebibliography}
\end{document}